\documentclass[10pt,letterpaper]{article}
\usepackage{opex3}
\usepackage{cite}
\usepackage{amsmath}
\usepackage{amssymb}
\usepackage{setspace}
\begin{document}

\title{Stimulated brillouin scattering in nanoscale silicon step-index waveguides: \\
A general framework of selection rules and calculating SBS gain}
\author{Wenjun Qiu,$^1$ Peter T. Rakich,$^{2,3}$ Marin Solja\v{c}i\'{c},$^1$ and Zheng Wang$^{4*}$}
\address{
$^1$Department of Physics, Massachusetts Institute of Technology, Cambridge, MA 02139 USA \\
$^2$Sandia National Laboratories, PO Box 5800 Albuquerque, NM 87185 USA\\
$^3$Department of Applied Physics, Yale University, New Haven, CT 06520 USA\\
$^4$Department of Electrical and Computer Engineering, University of Texas at Austin, Austin, TX 78758 USA}
\email{$^*$zheng.wang@austin.utexas.edu}

\begin{abstract}
We develop a general framework of evaluating the gain coefficient of Stimulated Brillouin Scattering (SBS) in optical waveguides via the overlap integral between optical and elastic eigen-modes. We show that spatial symmetry of the optical force dictates the selection rules of the excitable elastic modes. By applying this method to a rectangular silicon  waveguide, we demonstrate the spatial distributions of optical force and elastic eigen-modes jointly determine the magnitude and scaling of SBS gain coefficient in both forward and backward SBS processes. We further apply this method to inter-modal SBS process, and demonstrate that the coupling between distinct optical modes are necessary to excite elastic modes with all possible symmetries.
\end{abstract}


\bibliographystyle{osajnl}
\bibliography{phononics}


\section{Introduction}

Stimulated Brillouin Scattering (SBS) is a third-order nonlinear process with a broad range of implications in efficient phonon generation \cite{Chiao64,Russell06}, optical frequency conversion \cite{Kobyakov10,Russell09,Pant11}, slow light \cite{Song05,Song06,Gaeta05,Pant08}, and signal processing  \cite{Carmon09,Gauthier07}. The SBS process, measured by the coupling between optical waves and elastic waves, is recently discovered to be enhanced by orders of magnitude in nanoscale optical waveguides \cite{PRX}. Since the transverse dimensions of a nanoscale waveguide are close to or smaller than the wavelengths of optical and elastic waves, both waves are strongly confined as discrete sets of eigenmodes. Particularly strong SBS occurs, when two optical eigenmodes resonantly couple to an elastic eigenmode \cite{Boyd,Agrawal}. In general, the interference of pump and Stokes waves generates a time-varying and spatially-dependent optical force. On resonance, the optical force is simultaneously frequency-matched and phase-matched to an elastic mode, and results in strong mechanical vibration in the waveguide. The associated deformation is unusually large for nanoscale waveguides, because of the contribution from the surface forces and the large surface area. Such deformation in turn leads to highly efficient scattering between the pump and Stokes photons. However, because both transverse and longitudinal waves exist in elastic waves, together with the depolarization of elastic waves at the material boundaries \cite{Royer1}, a large number of elastic eigenmodes with disparate spatial profiles can be involved. It is therefore crucial to develop a theoretical framework that links the excitation of individual elastic modes with the properties of pump and Stokes waves. On one hand, this framework elucidates the contributions from individual elastic modes towards the overall SBS nonlinearity, thereby pointing towards designing traveling-wave structures that deliberately enhance or suppress SBS nonlinearity. On the other hand, this knowledge also enables one to devise optical fields that target the generation of specific phonon modes, in the context of efficiently transducing coherent signals between optical and acoustic domains.  

Generally, the strength of SBS nonlinearity is characterized by the SBS gain. This coefficient, in the past, has been theoretically derived from various forms of overlap integral between optical waves and elastic waves \cite{Boyd,Agrawal,Bahloul95,Russell09,Gauthier11,McCurdy05,Huang11,Russell10,Russell11}. While accurate for waveguides larger than a few microns, these treatments underestimate the SBS gain by orders of magnitude for nanoscale waveguides \cite{PRX}, for a couple of reasons. First, conventional treatments are based on nonlinear polarization current, and the associated electrostriction \emph{body} forces. The calculated SBS gain fails to capture boundary nonlinearities such as electrostriction pressure and radiation pressure at the waveguide surfaces. The latter two nonlinearities become significant, and in some cases, dominate in nanoscale waveguides, where the relative surface area is much larger than that of the microscale waveguides. Second, most previous studies assume the optical modes are \emph{linearly} polarized, or simplify the elastic modes with a \emph{scalar} density wave. For nanoscale waveguides, the vectorial nature and the nontrivial spatial distribution of both optical and elastic eigenmodes have to be fully considered.

In this article, we present a general method of calculating SBS gains via the overlap integral between optical forces and elastic eigen-modes. Within this formalism, all three types of optical forces are taken into account: the bulk and boundary nonlinearities are formulated as bulk and boundary integrals over the waveguide cross-section. In addition, both the optical and elastic modes are treated as vector fields, allowing for the most general forms of dielectric and elastic tensors, both forward and backward launching conditions, as well as intra-modal and inter-modal couplings. Armed with this formalism, we study the SBS process of a rectangular silicon waveguide. We show that all the optical forces in the forward SBS configuration are transverse. The constructive combination of electrostriction force and radiation pressure occurs for certain elastic modes with matching symmetries, and results in large forward SBS gains. In contrast, the optical forces in the backward SBS configuration are largely longitudinal, and the maximal backward SBS gain among all the elastic modes approaches the gain coefficient predicted by conventional SBS theory. We further apply this formalism to inter-modal SBS: by coupling optical modes with distinct spatial symmetries, optical forces with all possible symmetries can be generated, which offers a great deal of flexibility in producing elastic modes with a wide range of spatial symmetries.

\section{Calculating the SBS gain via overlap integral}

We start with a general framework of calculating the SBS gain from the field profiles of both the optical and elastic eigen-modes of a waveguide. The axial direction of the axially invariant waveguide is designated as the $x$ direction. In a typical SBS process, a pump wave $\mathbf{E}_p e^{i(k_p x - \omega_p t)}$ and a Stokes wave $\mathbf{E}_s e^{i(k_s x - \omega_s t)}$ generate dynamic optical forces that vary in space with a wavevector $q = k_p - k_s$, and oscillate in time at the beat frequency $\Omega = \omega_p - \omega_s$. 

Depending on the launching conditions, SBS can be categorized into forward SBS (FSBS) and backward SBS (BSBS). In FSBS, the pump and Stokes waves are launched in the same direction, generating axially-invariant optical forces, which excite standing-wave elastic modes \cite{Russell09}. In BSBS, the pump and Stokes waves propagate along opposite directions, generating axially-varying optical forces, which excite traveling-wave elastic modes. Besides launching the pump and Stokes waves into the same spatial optical mode of the waveguide, SBS can also occur with the pump and Stokes waves belonging to disparate spatial modes, for example, by launching into modes with different polarizations \cite{Russell10}. Such inter-modal SBS are important for optical signal isolation \cite{Russell10,Yu08,Huang11, Russell11} and Brillouin cooling of mechanical devices \cite{Carmon12}. These different launching conditions will be individually addressed in the later part of the article.

The optical forces that mediate SBS includes the well-known electrostriction force\cite{Carmon12, Wang10}, and radiation pressure whose contribution is only recently recognized\cite{PRX}. Electrostriction is an intrinsic material nonlinearity, which arises from the tendency of materials to become compressed in regions of high optical intensity. Conventionally, only the electrostriction in the form of a body force is considered as the dominant component \cite{Boyd,Agrawal}. However, the discontinuities in both optical intensities and photoelastic constants generates electrostriction pressure on material boundaries, abundant in nanostructures. Radiation pressure is another boundary nonlinearity, arising from the momentum exchange of light with the material boundaries with discontinuous dielectric constant \cite{Povinelli05,Wang11}. Radiation pressure is also radically enhanced in nanoscale structures, exemplified in a wide variety of optomechanics applications \cite{Vahala08,Painter07,Painter09,Rakich07,Tang09,Baets09}. In this formalism, by considering the superposition of all three forms of optical forces, not only can the SBS gain be more accurately evaluated for nanoscale waveguides, one can also take advantage of the coherent interference between these three components, to gain new degree of freedoms of tailoring SBS process.

This total optical force, i. e. the coherent superposition of all three components mentioned above, can excite mechanical vibrations which enables the parametric conversion between pump and Stokes waves. This process can be describe by the following relation \cite{Boyd}:
\begin{equation}
\frac{dP_s}{dx} = gP_pP_s - \alpha_s P_s
\end{equation}
Here, $P_p$ and $P_s$ are the guided power of the pump and Stokes waves, and $g$ is the SBS gain. Through particle flux conservation, SBS gain is given by the following formula \cite{PRX}:
\begin{equation}
g(\Omega) = \frac{\omega_s}{2\Omega P_p P_s} Re\left\langle \mathbf{f}, \frac{d\mathbf{u}}{dt} \right\rangle 
\end{equation}
where $\mathbf{f}$ is the total optical force generated by pump and Stokes waves, and $\mathbf{u}$ describes the elastic deformation of the waveguide induced by $\mathbf{f}$. The inner product between two vector fields is defined as the overlap integral over the waveguide cross-section:
\begin{equation}
\langle \mathbf{A},\mathbf{B} \rangle \triangleq
\int \mathbf{A}^* \cdot \mathbf{B} \ dydz
\end{equation}
The optical power of a waveguide is given by $P=v_g \langle \mathbf{E}, \epsilon \mathbf{E} \rangle /2$, where $v_g$ is the optical group velocity. Therefore,
\begin{equation}
g(\Omega) = \frac{2\omega_s}{v_{gp}v_{gs}}
\frac{Im \langle \mathbf{f},\mathbf{u} \rangle
\label{G1}}
{\langle \mathbf{E}_p,\epsilon\mathbf{E}_p \rangle
\langle \mathbf{E}_s,\epsilon\mathbf{E}_s \rangle}
\end{equation}

To further simply Eq. (\ref{G1}), we consider the equation governing the elastic response $\mathbf{u}e^{-i\Omega t}$ under external forces $\mathbf{f}e^{-i\Omega t}$. We begin with the ideal case, neglecting the mechanical loss  \cite{Royer1}:
\begin{equation}
-\rho \Omega^2 u_i =  \frac{\partial}{\partial x_j} c_{ijkl} \frac{\partial u_l}{\partial x_k} + f_i
\end{equation}
Here $\rho$ is the mass density, and $c_{ijkl}$ is the elastic tensor. $c_{ijkl}$ has two important properties: it is symmetric with respect to the first two and last two indices ($c_{ijkl}=c_{jikl}$, $c_{ijlk}=c_{ijkl}$); the interchange of the first two indices and the last two does not affect the value of $c_{ijkl}$: $c_{klij}=c_{ijkl}$ \cite{Royer1}. In the absence of a driving force $\mathbf{f}$, the equation above becomes the eigen-equation of elastic waves. Using the symmetry properties of $c_{ijkl}$, we can show that the operator in the left hand side of the eigen-equation is Hermitian. Therefore, the eigen-mode $\mathbf{u}_m e^{-i\Omega_m t}$ satisfies orthogonality condition:
\begin{equation}
\langle \mathbf{u}_m, \rho \mathbf{u}_n \rangle = 
\delta_{mn} \langle \mathbf{u}_m, \rho \mathbf{u}_m \rangle
\label{on}
\end{equation}
When $\mathbf{f}$ is present, $\mathbf{u}$ can be decomposed in terms of eigen-modes $\mathbf{u} = \sum_m b_m \mathbf{u}_m$. Using the orthogonality condition, we have:
\begin{equation}
b_m = \frac{\langle \mathbf{u}_m, \mathbf{f} \rangle}
{\langle \mathbf{u}_m,\rho \mathbf{u}_m \rangle}
\frac{1}{\Omega_m^2 - \Omega^2}
\end{equation}
We now consider the more general and practical cases, where mechanical loss is present. The commonly encountered mechanical loss mechanisms are air damping, thermoelastic dissipation, and clamping losses \cite{Kenny08}. The first-order effect of loss can be captured by changing $\Omega_m$ to a complex value, $\Omega_m - i \Gamma_m/2$. Assuming the mechanical quality factor $Q_m=\Omega_m / \Gamma_m$ is well above 1, we have,
\begin{equation}
b_m = \frac{\langle \mathbf{u}_m, \mathbf{f} \rangle}
{\langle \mathbf{u}_m,\rho \mathbf{u}_m \rangle}
\frac{1}{\Omega_m \Gamma_m}
\frac{\Gamma_m/2}{\Omega_m - \Omega - i\Gamma_m/2}
\label{b_m}
\end{equation}
Inserting Eq. (\ref{b_m}) into Eq. (\ref{G1}), we can see that the total SBS gain is the sum of SBS gains of individual elastic modes.
\begin{equation}
g(\Omega) = \sum_m G_m \frac{(\Gamma_m/2)^2}{(\Omega - \Omega_m)^2 + (\Gamma_m/2)^2}
\label{g}
\end{equation}
The SBS gain of a single elastic mode has a Lorentian shape and a peak value:
\begin{equation}
G_m = \frac{2 \omega}{\Omega_m \Gamma_m v_{gp} v_{gs}}
\frac{|\langle \mathbf{f},\mathbf{u}_m \rangle|^2}
{\langle \mathbf{E}_p,\epsilon\mathbf{E}_p \rangle
\langle \mathbf{E}_s,\epsilon\mathbf{E}_s \rangle
\langle \mathbf{u}_m,\rho \mathbf{u}_m \rangle}
\label{G2}
\end{equation}
where we have used the fact that $\Omega \ll \omega_p, \omega_s$ and $\omega_p \approx \omega_s = \omega$.

Equation (\ref{G2}) provides a general method to calculate the SBS gain of a waveguide with arbitrary cross-section. For example, with the finite element method, one can numerically calculate the pump and Stokes optical modes at a given $\omega$ and the elastic modes at the phase-matching wavevector $q = k_p - k_s$. The SBS of each elastic mode can then be calculated by taking the overlap integral between the derived optical forces and the elastic displacement. Here, body forces are integrated over the waveguide cross-section, while pressures are integrated over the waveguide boundaries. Overall, Eq. (\ref{G2}) shows that the SBS gain is determined by the frequency ratio, the mechanical loss factor, the optical group velocities, and the overlap integral between optical forces and elastic eigen-modes. In addition, Eq. (\ref{G2}) provides a convenient way to separate the effects of various optical forces. Specifically, the overlap integral is the linear sum of all optical forces:
\begin{equation}
\langle \mathbf{f},\mathbf{u}_m \rangle = \sum_n \langle \mathbf{f}_n,\mathbf{u}_m \rangle
\end{equation}  
The amplitudes of individual overlap integrals determine the maximal potential contribution from each form of optical forces, while their relative phases produce the interference effect.

A key step of applying Eq. (\ref{G2}) is to calculate optical forces from pump and Stokes waves. Electrostriction forces are derived from electrostriction tensor. The instantaneous electrostriction tensor is given by:
\begin{equation}
\sigma_{ij} = - \frac{1}{2}\epsilon_0 n^4 p_{ijkl} E_k E_l
\label{sigma1}
\end{equation}
where $n$ is the refractive index, and $p_{ijkl}$ is the photoelastic tensor \cite{Royer2}. In a waveguide system, the total electric field is given by $(\mathbf{E}_p e^{i(k_p x - \omega_p t)} + \mathbf{E}_s e^{i(k_s x - \omega_s t)})/2 + c.c$. Inserting this expression to Eq. (\ref{sigma1}), and filtering out the components with frequency $\Omega$, we arrive at the time-harmonic electrostriction tensor $\sigma_{ij} e^{i(q x - \Omega t)}$:
\begin{equation}
\sigma_{ij} = - \frac{1}{4}\epsilon_0 n^4 p_{ijkl} (E_{pk}E_{sl}^* + E_{pl}E_{sk}^*)
\label{sigma2}
\end{equation}
Since common materials used in integrated photonics have either cubic crystalline lattice (e.g. silicon) or are isotropic (e.g. silica glass), and most waveguide structures are fabricated to be aligned with the principal axes of the material, we consider the crystal structure of the waveguide material to be symmetric with respect to planes $x=0$, $y=0$, and $z=0$. Therefore, $p_{ijkl}$ is zero if it contains odd number of a certain component. In the contracted notation, Eq. (\ref{sigma2}) can be written as:
\begin{equation}
\begin{bmatrix}
\sigma_{xx} \\
\sigma_{yy} \\
\sigma_{zz} \\
\sigma_{yz} \\
\sigma_{xz} \\
\sigma_{xy} \\ 
\end{bmatrix} = -\frac{1}{2}\epsilon_0 n^4
\begin{bmatrix}
p_{11} & p_{12} & p_{13} & & & \\
p_{12} & p_{22} & p_{23} & & & \\
p_{13} & p_{23} & p_{33} & & & \\
& & & p_{44} & & \\ 
& & & & p_{55} & \\
& & & & & p_{66} \\
\end{bmatrix}
\begin{bmatrix}
E_{px}E_{sx}^* \\
E_{py}E_{sy}^* \\
E_{pz}E_{sz}^* \\
E_{py}E_{sz}^* + E_{pz}E_{sy}^* \\
E_{px}E_{sz}^* + E_{pz}E_{sx}^* \\
E_{px}E_{sy}^* + E_{py}E_{sx}^* \\
\end{bmatrix}
\label{sigma3}
\end{equation}
Electrostriction force is given by the divergence of electrostriction tensor. In a system consisting of domains of  homogeneous materials, electrostriction forces can exist inside each material (electrostriction body force) and on the interfaces (electrostriction pressure). Electrostriction body force is expressed as $\mathbf{f}^{ES} e^{i(q x - \Omega t)}$:
\begin{eqnarray}
f_x^{ES} &=& - iq\sigma_{xx} - \partial_y \sigma_{xy} - \partial_z \sigma_{xz} \nonumber \\
f_y^{ES} &=& - iq\sigma_{xy} - \partial_y \sigma_{yy} - \partial_z \sigma_{yz} \label{f} \\
f_z^{ES} &=& - iq\sigma_{xz} - \partial_y \sigma_{zy} - \partial_z \sigma_{zz} \nonumber
\end{eqnarray}
Electrostriction pressure on the interface between material 1 and 2 is given by $\mathbf{F}^{ES}e^{i(qx - \Omega t)}$ (normal vector $n$ points from 1 to 2):
\begin{equation}
F_i^{ES} = (\sigma_{1ij} - \sigma_{2ij}) n_j
\label{FES}
\end{equation}
With a particular choice of phase, an optical mode of the waveguide, $\mathbf{E}e^{i(kx - \omega t)}$, can be expressed as an imaginary-valued $E_x$ and real-valued $E_{y}$, $E_{z}$. From Eq. (\ref{sigma3}), we can see that $\sigma_{xx}$, $\sigma_{yy}$, $\sigma_{zz}$, and $\sigma_{yz}$ are real while $\sigma_{xy}$ and $\sigma_{xz}$ are imaginary. From Eq. (\ref{f}) and Eq. (\ref{FES}), we can also see that for both electrostriction body force and electrostriction pressure, the transverse component is real while the longitudinal component is imaginary.

Radiation pressure is derived from Maxwell Stress Tensor (MST). For a dielectric system ($\mu=1$) without free charges ($\rho=0,J=0$), radiation pressure is localized where the gradient of $\epsilon$ is nonzero \cite{Gordan73,Johnson02}. For a system consisting of homogeneous materials, radiation pressure only exists on the interfaces. The electric part of instantaneous MST is:
\begin{equation}
T_{ij} = \epsilon_0 \epsilon (E_i E_j - \frac{1}{2}\delta_{ij}E^2)
\label{T}
\end{equation}
The instantaneous pressure on the interface between material 1 and 2 is:
\begin{equation}
F^{RP}_i = (T_{2ij} - T_{1ij})n_j
\end{equation}
By decomposing the electric field into its normal and tangential components with respect to the dielectric interface $\mathbf{E} = E_n\mathbf{n} + E_t\mathbf{t}$, and using the boundary condition $\epsilon_1 E_{1n} = \epsilon_2 E_{2n} = D_n$ and $E_{1t} = E_{2t} = E_t$, we can show that:
\begin{equation}
\mathbf{F}^{RP} = -\frac{1}{2}\epsilon_0 E_t^2 (\epsilon_2 - \epsilon_1) \mathbf{n}
+ \frac{1}{2}\epsilon_0^{-1} D_n^2 (\epsilon_2^{-1} - \epsilon_1^{-1}) \mathbf{n}
\end{equation}
Inserting the total electric field $(\mathbf{E}_p e^{i(k_p x - \omega_p t)} + \mathbf{E}_s e^{i(k_s x - \omega_s t)})/2 + c.c$ to the expression above, and filtering out the components with frequency $\Omega$, we can get the time-harmonic radiation pressure $\mathbf{F}^{RP} e^{i(q x - \Omega t)}$:
\begin{equation}
\mathbf{F}^{RP} = -\frac{1}{2}\epsilon_0 E_{pt}E_{st}^* (\epsilon_2 - \epsilon_1) \mathbf{n}
+ \frac{1}{2}\epsilon_0^{-1} D_{pn}D_{sn}^* (\epsilon_2^{-1} - \epsilon_1^{-1}) \mathbf{n}
\label{FRP}
\end{equation}
Equation (\ref{FRP}) shows that radiation pressure is always in the normal direction. For axially invariant waveguide, this also means radiation pressure is transverse and real.

Combining Eq. (\ref{G2}) with the calculation of optical forces, we are ready to numerically explore the SBS nonlinearity of nanoscale waveguides. Before that, it is instructive to compare Eq. (\ref{G2}) with the conventional BSBS gain \cite{Agrawal}. We can show that Eq. (\ref{G2}) converges to the conventional BSBS gain under the plane-wave approximation for both optical and elastic modes. Specifically, consider the coupling between two counter propagating optical plane-waves through an elastic plane-wave. The optical plane-wave is linearly polarized in $y$ direction. The elastic plane-wave is purely longitudinal traveling at velocity $V_L$. Under this setup, nonzero optical forces include the longitudinal electrostriction body force, and the transverse components of electrostriction pressure and radiation pressure. Only the longitudinal electrostriction body force contributes nonzero overlap integral:
\begin{equation}
f^{ES}_x = -iq\sigma_{xx} = \frac{1}{2}iq\epsilon_0 n^4 p_{12} E_y ^2 
\end{equation}
Inserting this expression into Eq. (\ref{G2}), and using the fact that $\Omega = q V_L$ and $q = 2k$, we arrive at:
\begin{equation}
G_0 = \frac{\omega^2 n^7 p_{12}^2}{c^3 \rho V_L \Gamma} \frac{1}{A}
\label{G0}
\end{equation}
where $A$ is the cross-sectional area of the waveguide. This is exactly the conventional BSBS gain. For waveguides with transverse dimension much greater than the free-space wavelength of light, the plane-wave approximation is valid, and Eq. (\ref{G2}) converges to $G_0$. For nanoscale waveguides, Eq. (\ref{G2}) can deviate from $G_0$ significantly because of the vectorial nature of optical and elastic modes, nontrivial mode profiles, as well as the enhanced boundary nonlinearities.

\section{Rectangular silicon waveguide: intra-modal coupling}

\begin{figure}[p]
\centering\includegraphics[width=\textwidth]{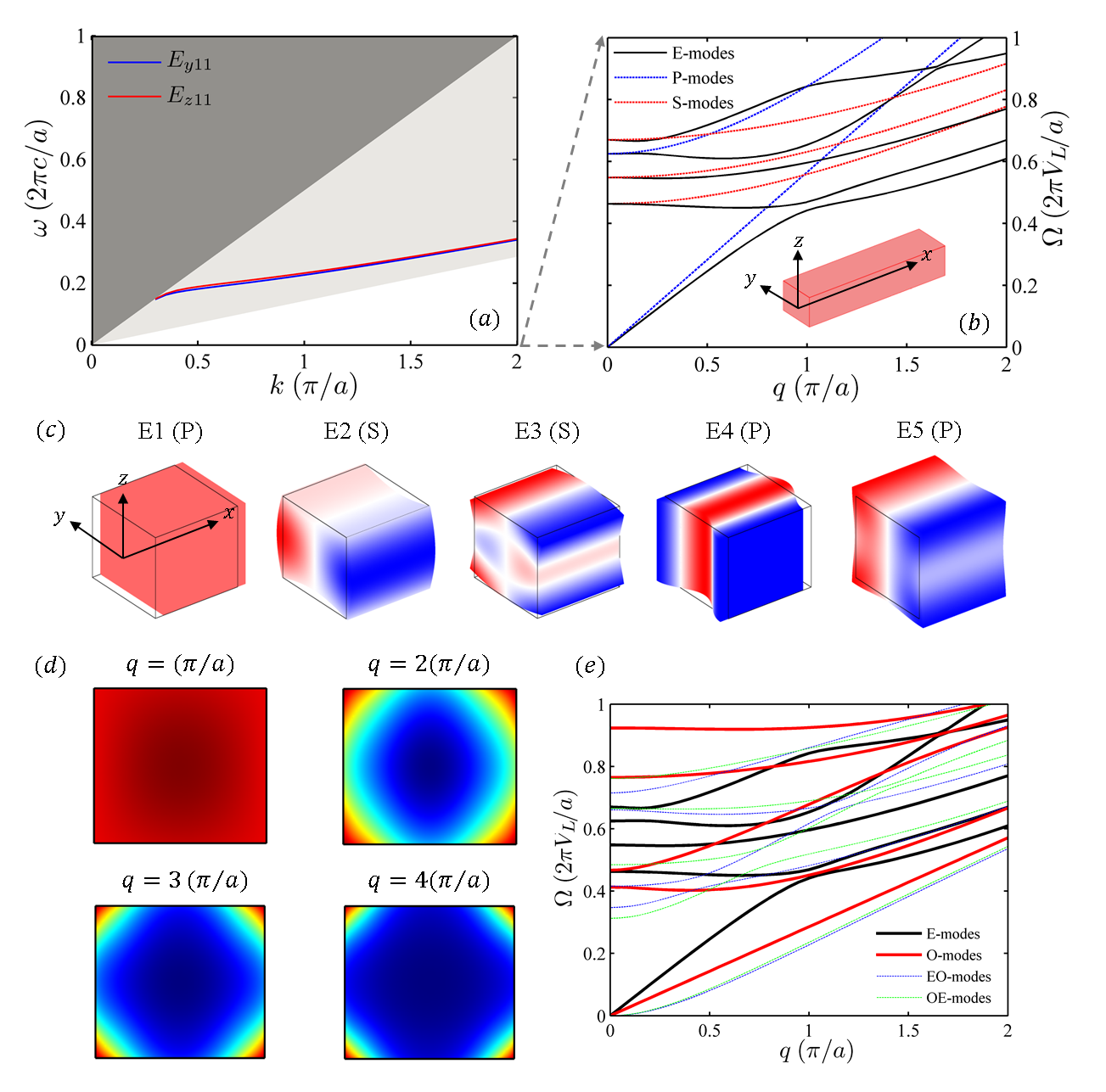}
\caption{The guided optical and elastic modes of a rectangular silicon waveguide. Optical frequency is in unit of $2 \pi c / a$, while elastic frequency is in unit of $2 \pi V_L / a$. $V_L = \sqrt{E/\rho} = 8.54 \times 10^3$m/s is the velocity of longitudinal elastic waves in bulk silicon. (a) Dispersion relation of optical modes $E_{y11}$ and $E_{z11}$. (b) Dispersion relation of elastic modes which have even symmetry with respect to planes $y=0$ and $z=0$. E-modes (black lines) are the actual eigen-modes of the silicon waveguide, with silicon-air interfaces treated as free boundaries. For comparison, the dispersion relations of purely longitudinal modes (designated as P-modes, blue curves) and purely transverse modes (designated as S-modes, red curves) are included. They are constrained respectively with x-only displacement, and y-z-only movements. At $q=0$, E-modes manifest as either P-modes or S-modes. (c) The displacement profiles of mode E1 through E5 at $q=0$, with the peak deformation shown. The color represents y-displacement ($u_y$) for S-type E-modes and x-displacement ($u_x$) for P-type E-modes. Blue, white, and red correspond to negative, zero, and positive values respectively. Mode E1 experiences a DC longitudinal offset at $\Omega=0$. (d) The evolution of mode E1 as $q$ increases. The color-map corresponds to the amplitude of displacement vector $|u|^2$ with blue and red corresponding to zero and maximal values. (e) The dispersion relations of O-modes (odd about both symmetry planes), EO-modes (even about $y=0$ and odd about $z=0$), and OE-modes (odd about $y=0$ and even about $z=0$).}
\label{fig1}
\end{figure}

Intra-modal process is concerned with the configuration where the pump and the Stokes waves are launched into the same spatial optical mode of the waveguide. In this section, we apply the general formalism to study the intra-modal SBS process of a silicon waveguide suspended in air. Silicon waveguides are of particular interest, because they can be fabricated from standard SOI platforms. A suspended silicon waveguide can provide tight optical confinement through the large refractive index contrast and nearly perfect elastic confinement through the dramatic impedance mismatch with air. In addition, since radiation pressure is proportional to the difference of dielectric constants across waveguide boundaries and electrostriction force is quadratic over refractive index, both kinds of optical forces are significantly enhanced in high index contrast structures such as silicon waveguides. Here, we consider a silicon waveguide with a rectangular cross-section of $a$ by $0.9a$ (Fig. \ref{fig1}(a) insert). For silicon, we use refractive index $n=3.5$, Young's modulus $E=170\times 10^9$ Pa, Poisson's ratio $\nu=0.28$, and density $\rho=2329$kg/m\textsuperscript{2}. In addition, we assume that the [100], [010], and [001] symmetry direction of this crystalline silicon coincide with the $x$, $y$, and $z$ axis respectively. Under this orientation, the photo-elastic tensor $p_{ijkl}$ in the contracted notation is $[p_{11},p_{12},p_{44}] = [-0.09,0.017,-0.051]$ \cite{Briddon06}. The structure has two symmetry planes $y=0$ and $z=0$. Both optical modes and elastic modes are either even or odd with respect to these planes.

We categorize the fundamental optical modes in the two polarizations as $E_{y11}$ and $E_{z11}$ (Fig. \ref{fig1}(a)). $E_{y11}$ is even with respect to plane $z=0$ and odd with respect to plane $y=0$ with a large $E_y$ component. $E_{z11}$ has the opposite symmetries and slightly higher frequencies. We normalize the angular frequency $\omega$ in unit of $2\pi c/a$. Throughout the study, we assume the pump wavelength at 1.55$\mu$m. Therefore, a different normalized frequency along the optical dispersion relation implies a different $a$. For intra-modal coupling, we assume that pump and Stokes waves come from $E_{y11}$. Since $\Omega/\omega \approx V_L / c $ is on the order of $10^{-4}$, pump and Stokes waves approximately corresponds to the same waveguide mode $\mathbf{E}e^{i(k x - \omega t)}$. The induced optical force in intra-modal coupling is always symmetric with respect to planes $y=0$ and $z=0$. Therefore, we only need to consider elastic modes with the same spatial symmetry (Fig. \ref{fig2}(b)). Using a finite element solver, we calculate the eigen-mode of the suspended waveguide with free boundary conditions (E-modes). To illustrate the hybrid nature of E-modes, we also calculate purely longitudinal modes (P-modes) and purely transverse modes (S-modes) by forcing $u_{y,z}=0$ or $u_{x}=0$ throughout the waveguide. The dispersion relations indicates that E-modes are either P-mode or S-mode at $q=0$, but become a hybridized wave with both longitudinal and transverse components at $q \neq 0$. At $q=0$,  the mirror reflection symmetry with respect to plane $x = 0$ is conserved . Odd (even) modes with respect to plane $x=0$ are purely longitudinal (transverse), separating E-modes into P-modes and S-modes. At nonzero $q$, silicon-air boundaries hybridize the P-modes and the S-modes, resulting in E-modes with both longitudinal and transverse movement. Similar to the optical mode, we can choose a proper phase so that $u_x$ is imaginary while $u_{y,z}$ are real. Another observation is that the dispersion relation of mode E1 quickly deviates from that of mode P1 which is the longitudinal plane wave. The modal profiles at different $q$ indicates that mode E1 quickly evolves from a longitudinal plane wave to a surface-vibrating wave as $q$ increases (Fig. \ref{fig1}(d)).

\subsection{Forward SBS}

In traditional optical fibers, FSBS process is extremely weak, due to the excessively long wavelength and the vanishing frequency of the relevant elastic modes. However, waveguides with nanoscale feature sizes can efficiently produce FSBS, for example, in photonic crystal fibers \cite{Russell09} and suspended silicon waveguides \cite{PRX}. The frequency of the excitable elastic modes in FSBS is pinned by the structure, independent of the incident optical frequency. Both structures provide strong transverse phonon confinement, and these optical-phonon-like elastic modes are automatically phase-matched to higher orders of Stokes and anti-Stokes optical waves. The cascaded generation of such elastic modes through an optical frequency comb can enable efficient phonon generation with large quantum efficiency\cite{Russell09}.

\begin{figure}[p]
\centering\includegraphics[width=\textwidth]{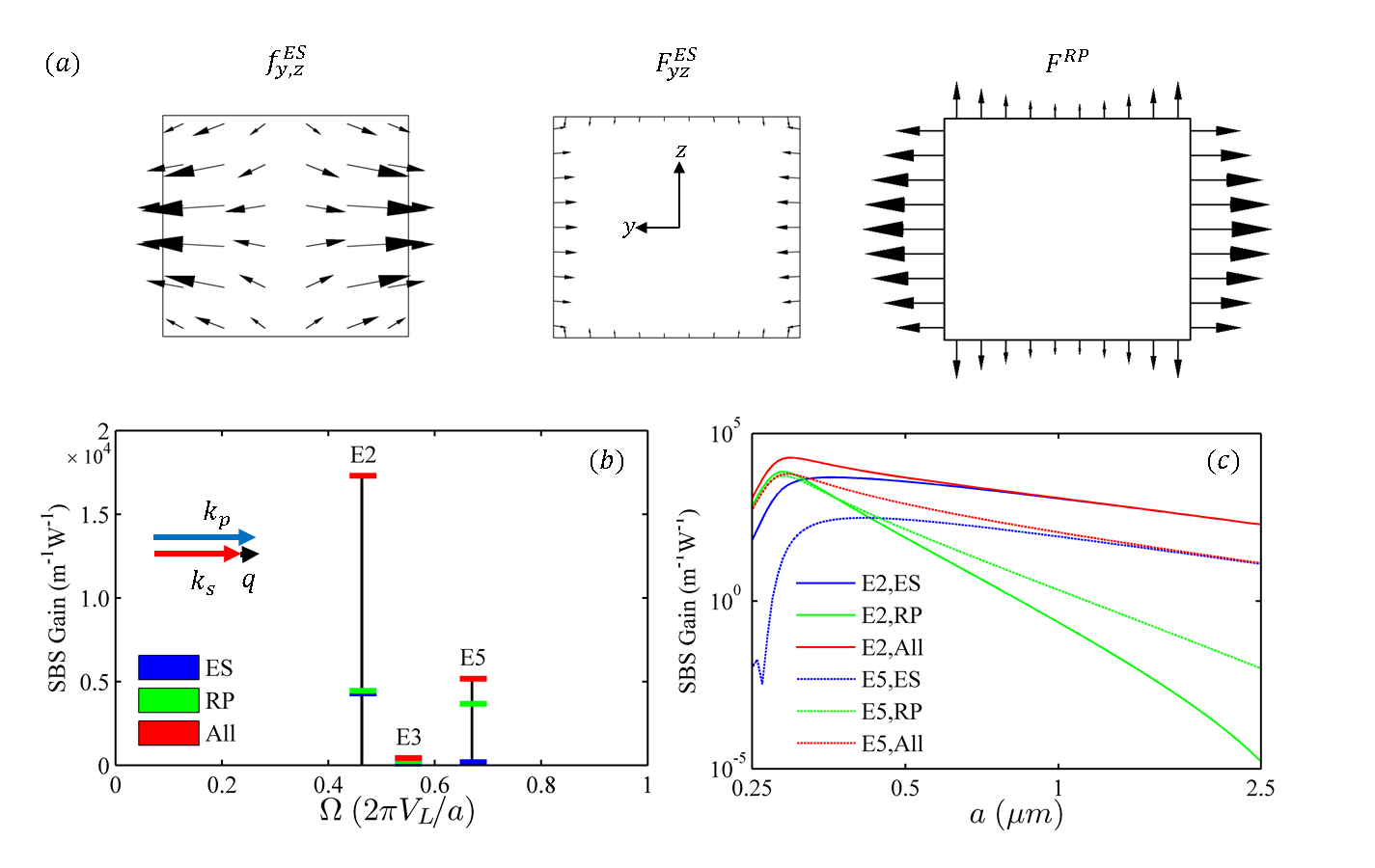}
\caption{Optical force distributions and the resultant gain coefficients of forward SBS process. In panels (a) and (b), the width of the waveguide is $a=315$nm, and the incident optical waves have $\omega=0.203(2\pi c/a)$, and $k=0.75(\pi/a)$. The elastic waves are generated at $q=0$. (a) The force distribution of electrostriction body force density, electrostriction surface pressure, and radiation pressure respectively. All three types of optical forces are transverse. (b) Calculated FSBS gains of the elastic modes, assuming a mechanical quality factor $Q_m=1000$. Blue, green, and red bars represent FSBS gains under three conditions: electrostriction-only, radiation-pressure-only, and the combined effects. Only the S-type E-modes have non-zero gains. (c) The scaling relation of FSBS gains as the device dimension $a$ is varied from 0.25$\mu$m to 2.5$\mu$m, color-coded similar to panel (b). Solid and dotted curves correspond to the gain coefficients for mode E2 and E5 respectively.}
\label{fig2}
\end{figure}

In FSBS, $\mathbf{E}_p = \mathbf{E}_s = \mathbf{E}$ and $q=0$. Equation (\ref{sigma3}) can be simplified to:
\begin{equation}
\begin{bmatrix}
\sigma_{xx} \\
\sigma_{yy} \\
\sigma_{zz} \\
\sigma_{yz} \\
\sigma_{xz} \\
\sigma_{xy} \\ 
\end{bmatrix} = -\frac{1}{2}\epsilon_0 n^4
\begin{bmatrix}
p_{11} & p_{12} & p_{13} & & & \\
p_{12} & p_{22} & p_{23} & & & \\
p_{13} & p_{23} & p_{33} & & & \\
& & & p_{44} & & \\ 
& & & & p_{55} & \\
& & & & & p_{66} \\
\end{bmatrix}
\begin{bmatrix}
|E_x|^2 \\
|E_y|^2 \\
|E_z|^2 \\
2Re(E_y E_z^*) \\
0 \\
0 \\
\end{bmatrix}
\end{equation}
Apparently, $\sigma_{xy} = \sigma_{xz} = 0$. From Eq. (\ref{f}) and Eq. (\ref{FES}), we conclude that $f^{ES}_x = F^{ES}_x = 0$. So both electrostriction force and radiation pressure in FSBS are transverse. We pick an operating point at $\omega = 0.203(2\pi c/a)$, $k = 0.75(\pi /a)$ with $a = 315$nm, and compute the force distribution (Fig. \ref{fig2}(a)). Electrostriction body force is largely in the $y$ direction, because $E_y$ is the dominant component in electric field and $|p_{11}|$ is about five times larger than $|p_{12}|$. Electrostriction pressure points inwards, and radiation pressure points outwards. Radiation pressure is about five times greater than electrostriction pressure. The transverse nature of optical force combined with the fact that elastic modes are either P-modes or S-modes at $q=0$ indicates that only S-modes have nonzero FSBS gains. The corresponding FSBS gains are calculated using $Q_m=1000$ for all the elastic modes (Fig. \ref{fig2}(b)). As expected, only S-modes E2, E3, and E5 have nonzero gains. Mode E2 has the largest gain of $1.72\times 10^4$m\textsuperscript{-1}W\textsuperscript{-1}, which comes from a constructive combination of electrostriction effect ($0.42\times 10^4$m\textsuperscript{-1}W\textsuperscript{-1}) and radiation pressure effect ($0.44\times 10^4$m\textsuperscript{-1}W\textsuperscript{-1}). Mode E5 has a total gain of $0.51\times 10^4$m\textsuperscript{-1}W\textsuperscript{-1}, which mainly comes from radiation pressure ($0.36\times 10^4$m\textsuperscript{-1}W\textsuperscript{-1}).

To illustrate the interplay between electrostriction and radiation pressure, we scale the waveguide dimension $a$ from 250nm to 2.5$\mu$m by raising the operating point in the optical dispersion relation from $0.16(2\pi c/a)$ to $1.61(2\pi c/a)$, and compute the corresponding FSBS gains for mode E2 and E5 (Fig. \ref{fig2}(c)). For both E2 and E5, the electrostriction-only FSBS gain scales as $1/a^2$ for large $a$. This can be understood by a detailed analysis of Eq. (\ref{G2}). Under normalization condition $\langle \mathbf{E}, \epsilon \mathbf{E} \rangle = 1$, the electrostriction tensor scales as $1/a^2$. Since electrostriction force is essentially the divergence of electrostriction tensor, the total electrostriction force that apply to the right half of the waveguide scales as $1/a^3$. Under normalization condition $\langle \mathbf{u}_m, \rho \mathbf{u}_m \rangle = 1$, $\mathbf{u}_m$ scales as $1/a$. So the overlap integral scales as $1/a^2$. Under a fixed quality factor, the electrostriction-only FSBS gain scales as $1/a^2$. 

Unlike the electrostriction contributions that run parallel in different modes, the radiation-pressure-only FSBS gain scales as $1/a^6$ for mode E5 and $1/a^8$ for mode E2. This can also be understood from a breakdown of Eq. (\ref{G2}). Given the input power, the sum of average radiation pressure on the horizontal and vertical boundaries of the rectangular waveguide is proportional to $(n_g - n_p)/A$, where $n_g$ ($n_p$) is the group (phase) index, and $A$ is the waveguide cross-section \cite{Wang11}. When the waveguide scales up, $n_g-n_p$ shrinks as $1/A$. As a result, the sum of average radiation pressure scales as $1/a^4$, and the radiation-pressure-only FSBS gain should scale as $1/a^6$. For mode E2, however, radiation pressures on the horizontal and vertical boundaries generate overlap integrals with opposite signs. It is the difference rather than the sum between the horizontal and vertical radiation pressures that determines the scaling of the gain coefficient. A closer examination reveals that although the overlap integral from radiation pressure on the horizontal/vertical boundaries scales as $1/a^4$, the net overlap integral scales as $1/a^5$, resulting in the $1/a^8$ scaling of the radiation-pressure-only FSBS gain for mode E2.

\subsection{Backward SBS}

In traditional optical fibers, BSBS process is the qualitatively different from FSBS, as it is the only configuration that allows strong photon-phonon coupling. Recent studies have demonstrated on-chip BSBS on chalcogenide rib waveguide \cite{Pant11}. Compared to fiber-based BSBS, chip-based BSBS has much larger gain coefficient and requires much smaller interaction length, which enables a wide variety of chip scale applications such as tunable slow light \cite{Pant12a}, tunable microwave photonic filter \cite{Pant12b}, and stimulated Brillouin lasers \cite{Li12}. Unlike FSBS where elastic modes at $q=0$ are excited, BSBS generates elastic modes at $q=2k$. Elastic modes traveling at different $q$ can be excited by varying the incident optical frequency.

\begin{figure}[p]
\centering\includegraphics[width=\textwidth]{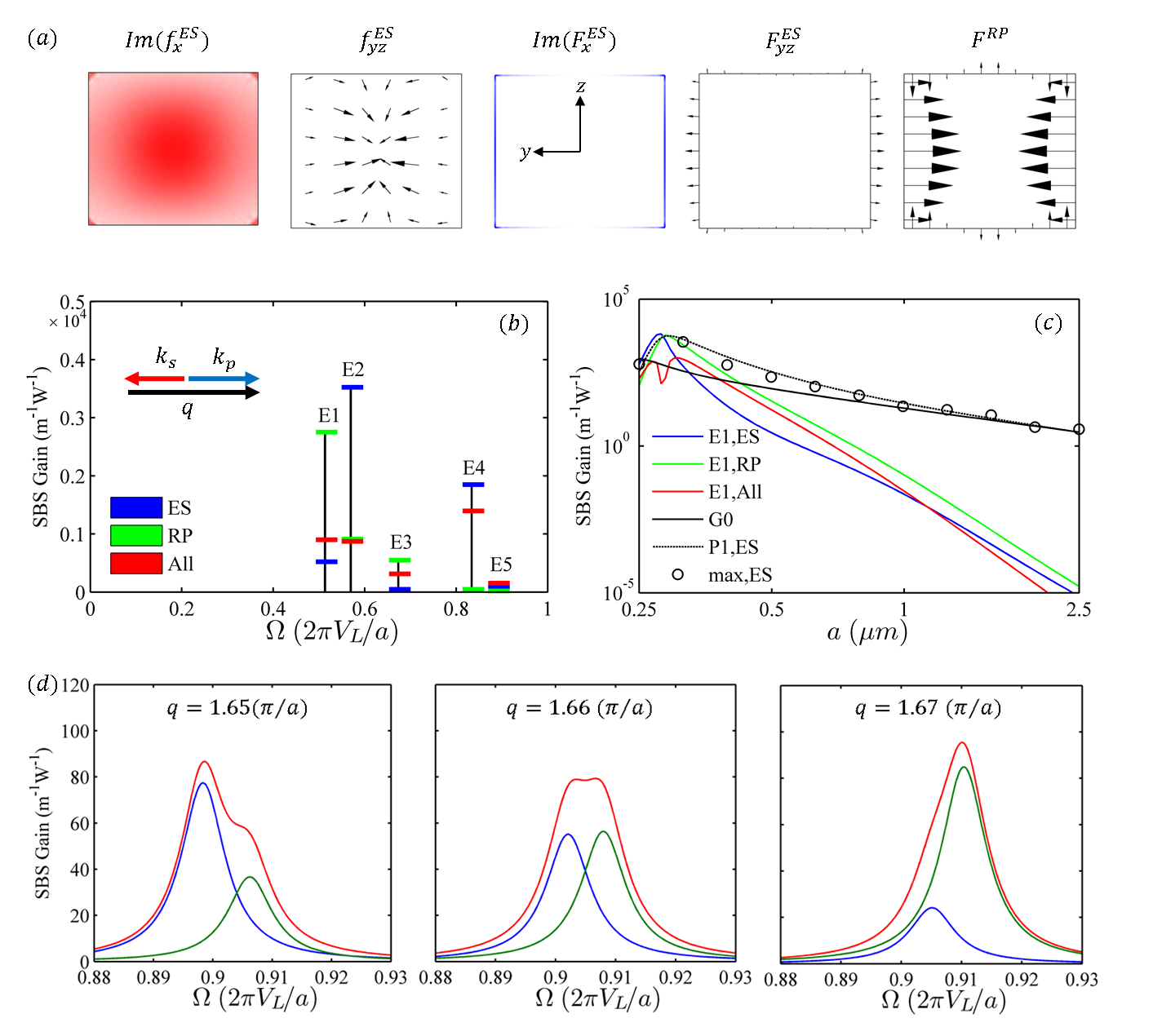}
\caption{Optical force distributions and the resultant gain coefficients of backward SBS process. In panels (a) and (b), the width of the waveguide is $a=315$nm, and the incident optical waves have $\omega=0.203(2\pi c/a)$, and $k=0.75(\pi/a)$. The elastic waves are generated at $q=1.5(\pi/a)$. (a) The force distribution of electrostriction body force density, electrostriction surface pressure, and radiation pressure respectively. Electrostriction have both longitudinal and transverse components. Radiation pressure are purely transverse. (b) Calculated BSBS gains of the elastic modes, assuming a mechanical quality factor $Q_m=1000$. Blue, green, and red bars represent BSBS gains under three conditions: electrostriction-only, radiation-pressure-only, and the combined effects.(c) The scaling relation of BSBS gains related to  mode E1 as $a$ is varied from 0.25$\mu$m to 2.5$\mu$m, color-coded similar to panel (b). For comparison, gain coefficients predicted by conventional SBS theory are shown as the solid black curve. The dotted black curve represents the electrostriction-only BSBS gain of the constrained mode P1. Black circles represent the largest electrostriction-only BSBS gain among all E-modes for a given $a$. (d) BSBS spectra near the anti-crossing between mode E4 and E5 around $q=1.66(\pi/a)$. The mechanical quality factor is assumed to be 100. The red lines represent the total BSBS gain. The blue and green lines represent contributions from mode E4 and E5 respectively.}
\label{fig3}
\end{figure}

In BSBS, $\mathbf{E}_p=\mathbf{E}$, $\mathbf{E}_s=\mathbf{E}^*$, and $q = 2k$. Equation (\ref{sigma3}) can be simplified to:
\begin{equation}
\begin{bmatrix}
\sigma_{xx} \\
\sigma_{yy} \\
\sigma_{zz} \\
\sigma_{yz} \\
\sigma_{xz} \\
\sigma_{xy} \\ 
\end{bmatrix} = -\frac{1}{2}\epsilon_0 n^4
\begin{bmatrix}
p_{11} & p_{12} & p_{13} & & & \\
p_{12} & p_{22} & p_{23} & & & \\
p_{13} & p_{23} & p_{33} & & & \\
& & & p_{44} & & \\ 
& & & & p_{55} & \\
& & & & & p_{66} \\
\end{bmatrix}
\begin{bmatrix}
E_x^2 \\
E_y^2 \\
E_z^2 \\
2E_yE_z \\
2E_xE_z \\
2E_xE_y \\
\end{bmatrix}
\end{equation}
All components of $\sigma_{ij}$ are nonzero, generating electrostriction force with both longitudinal and transverse components. We pick an operating point at $\omega = 0.203(2\pi c/a)$, $k = 0.75(\pi /a)$ with $a = 315$nm, and compute the force distribution (Fig. \ref{fig3}(a)). Electrostriction body force has large longitudinal component over the waveguide cross-section, which mainly comes from the $-iq\sigma_{xx}$ term in Eq. (\ref{f}). The hybrid nature of optical forces combined with the fact that all elastic modes are hybrid at nonzero $q$ indicates that all elastic modes have nonzero BSBS gains. We compute the corresponding BSBS gains using $Q_m=1000$ for all the elastic modes (Fig. \ref{fig3}(b)). For mode E1 and E2, electrostriction force and radiation pressure add up destructively, resulting in small BSBS gains of $0.089\times 10^4$m\textsuperscript{-1}W\textsuperscript{-1} and $0.086\times 10^4$m\textsuperscript{-1}W\textsuperscript{-1} respectively.

To study the evolution of elastic modes at different $q$ and its effect on BSBS gains, we vary $a$ from 250nm to 2.5$\mu$m and compute the corresponding BSBS gains for mode E1 (Fig. \ref{fig3}(c)). For comparison, we also compute the conventional BSBS gain $G_0$. The electrostriction-only BSBS gain of mode E1 decays very quickly. In contrast, $G_0$ scales as $1/a^2$ as required by Eq. (\ref{G0}). The reason is that, although mode E1 starts as a longitudinal plane wave for $q \approx 0$, it quickly evolves into surface-vibrating wave as $q$ increases (Fig. \ref{fig1}(d)). There are two ways to recover the scaling of $G_0$. First, we can force purely longitudinal movement by considering P-modes in Fig. \ref{fig1}(b). Mode P1 is exactly the longitudinal plane wave, characterized by uniform longitudinal vibrations across the waveguide cross-section and an approximately linear dispersion relation. The electrostriction-only BSBS for mode P1 does converge to $G_0$ (Fig. \ref{fig3}(c)). Second, the dispersion curve of mode P1 intersects with the dispersion curves of many E-modes as $q$ increases. For a given $q$, the E-modes which are close to the intersection point become P1-like with approximately uniform longitudinal vibrations across the waveguide cross-section. The electrostriction-only BSBS gain of these E-modes should be much larger than other E-modes, and close to that of mode P1. To verify this point, we compute the BSBS gains of a large number of E-modes. The maximal electrostriction-only BSBS gain among all the E-modes does converge to $G_0$ as $a$ exceeds several microns (Fig. \ref{fig3}(c)).

As mentioned above, the elastic dispersion relations can be fully explored by varying the operating point in the optical dispersion relation through phase-matching condition $q=2k$ in BSBS. One unique feature about the elastic dispersion relations is the abundance of anti-crossing between the hybridized elastic modes. The two elastic modes involved in anti-crossing typically have disparate spatial distributions and quite different BSBS gains. These two modes will exchange their spatial distributions and the corresponding BSBS gains when $q$ is scanned through the anti-cross region, as demonstrated in Fig. \ref{fig3}(d). Within the anti-crossing region, the spectrum of total SBS gain can have complicated shapes because of the overlap between modes with close eigen-frequencies. While the frequency response method in \cite{PRX} can only calculate the aggregated gain, the eigen-mode method developed here can not only separate the contributions from different elastic modes, but also parameterize the gain of individual modes with simple physical quantities. 

\section{Rectangular silicon waveguide: inter-modal coupling}

In inter-modal SBS, pump and Stokes waves belong to distinct optical modes. This feature can be exploited in several aspects. First, pump and Stokes waves can have orthogonal polarizations so that they can be easily separated with a polarizing beam splitter. Second, pump and Stokes waves can reside in optical modes with different dispersion relations. The nontrivial phase-matching condition can be exploited in optical signal isolation and Brillouin cooling of mechanical vibrations. More importantly, because the symmetry and spatial distribution of optical forces are jointly determined by pump and Stokes waves, in inter-modal SBS, the degree of freedoms of tailoring optical forces are essentially doubled, and the universe of excitable elastic modes is significantly expanded. For the rectangular waveguide discuss above, only elastic modes which are symmetric about planes $y=0$ and $z=0$ are excitable in intra-modal SBS. Elastic modes with all other symmetries can only be excited in inter-modal SBS, where the optical forces become anti-symmetric about a symmetry plane if pump and Stokes waves have opposite symmetries with respect to this plane.

\begin{figure}[p]
\centering\includegraphics[width=\textwidth]{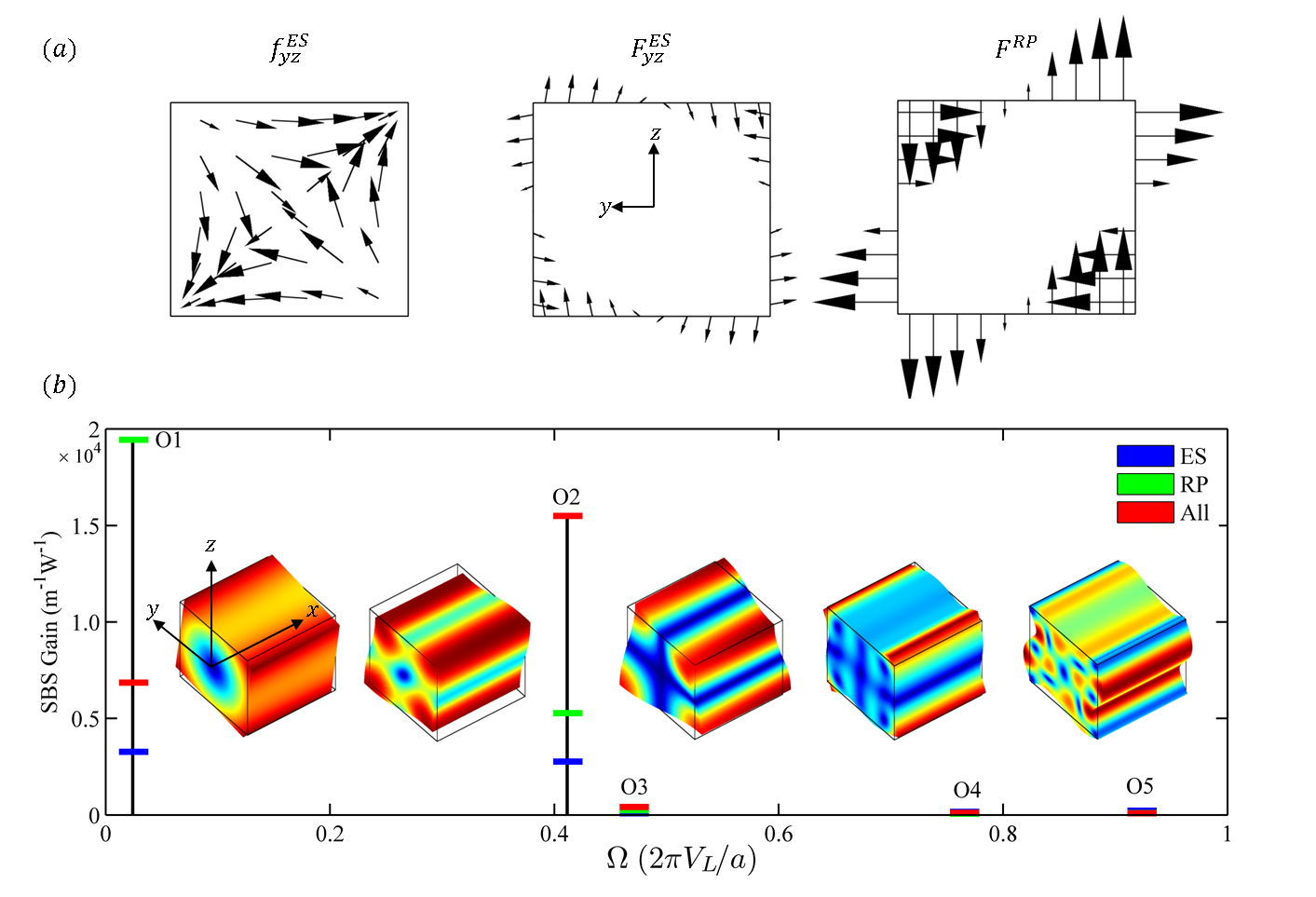}
\caption{Optical force distributions, relevant elastic modes, and the resultant gain coefficients of inter-modal SBS between $E_{y11}$ (pump) and $E_{z11}$ (Stokes). The width of the waveguide is set to be $a = 315$nm. The incident optical waves have $\omega = 0.203(2\pi c/a)$, with the pump-wave propagating at $k_p = 0.750(\pi/a)$, and the Stokes-wave propagating at $k_s = 0.665(\pi/a)$. The elastic waves are generated at $q = 0.085(\pi/a)$. (a) The force distribution of electrostriction body force density, electrostriction surface pressure, and radiation pressure respectively. The longitudinal forces (not shown here) are negligible, in comparison to the transverse forces. All optical forces are anti-symmetric with respect to planes $y = 0$ and $z = 0$, exciting elastic modes with the matching symmetry (designated as O-modes). (b) Calculated inter-modal SBS gains, assuming a mechanical quality factor $Q_m = 1000$. The insets illustrate the displacement profiles of mode O1 through O5 at $q = 0.085(\pi/a)$, at peak deformation. "Jet" colormap is used to shown the amplitude of \emph{total} displacement. Blue and red correspond to zero and maximum respectively.}
\label{fig4}
\end{figure}

For instance, we consider the coupling between $E_{y11}$ (pump) and $E_{z11}$ (Stokes). The operating point is $\omega = 0.203(2\pi c/a)$, $k_p = 0.750(\pi /a)$, $k_s = 0.665(\pi/a)$, and $q = 0.085(\pi/a)$ with $a = 315$nm. Because $E_{y11}$ and $E_{z11}$ have the opposite symmetries with respect to planes $y=0$ and $z=0$, the induced optical force is anti-symmetric with respect to both planes (Fig. \ref{fig4}(a)). Both electrostriction body force and radiation pressure try to pull the waveguide in one diagonal and squeeze the waveguide in the other diagonal. Electrostriction pressure has the opposite effect, but is much weaker than the radiation pressure. 

Under such optical force, elastic modes which are anti-symmetric with respect to planes $y=0$ and $z=0$ (O-modes) are excited. We calculate the SBS gains of mode O1 through O5 using $Q_m=1000$ for all the modes (Fig. \ref{fig4}(b)). Mode O1 represents a rotation around $x$ axis. The overlap integral is proportional to the torque. The $y$ component and $z$ component of the optical forces generate torques with opposite signs, which significantly reduces the total overlap integral. Mode O1 still has a sizable SBS gains because of its small elastic frequency $\Omega = 0.024(2\pi V_L/a)$. Mode O2 represents a breathing motion along the diagonal. Its modal profile coincides quite well with the optical force distribution. The constructive combination between electrostriction force and radiation pressure results in large gain coefficient of $1.54\times 10^4$m\textsuperscript{-1}W\textsuperscript{-1}. Mode O3 only have small gains since it is dominantly longitudinal while the optical forces are largely transverse. The SBS gains of mode O4, O5 and higher order modes are close to zero mainly because the complicated mode profiles is spatially mismatched with the optical force distribution: the rapid spatial oscillation of the elastic modes cancels out the overlap integrals to a large extent.

\section{Concluding remarks}

In this article, we present a general framework of calculating the SBS gain via the overlap integral between optical forces and elastic eigen-modes. Our method improved upon the frequency response representation of SBS gains \cite{PRX}. By decomposing the frequency response into elastic eigen-modes, we show that the SBS gain is the sum of many Lorentian components which center at elastic eigen-frequencies. The SBS gain spectrum is completely determined by the quality factor and maximal gain of individual elastic modes. Therefore, our method is conceptually clearer and computationally more efficient than the frequency response method. Through the study of a silicon waveguide, we demonstrate that our method can be applied to both FSBS and BSBS, both intra-modal and inter-modal coupling, both nanoscale and microscale waveguides. Both analytical expressions and numerical examples show that SBS nonlinearity is tightly connected to the symmetry, polarization, and spatial distributions of optical and elastic modes. The overlap integral formula of SBS gains provides the guidelines of tailoring and optimizing SBS nonlinearity through material selection and structural design.

\newpage


\end{document}